\pdfoutput=1
\documentclass[prb,aps,twocolumn]{revtex4-1}
\usepackage{amsmath,amssymb}
\usepackage{bm}
\usepackage{color}
\usepackage{comment}
\usepackage[svgnames]{xcolor}
\usepackage[colorlinks]{hyperref}
\hypersetup{linkcolor=DarkBlue,citecolor=DarkBlue,filecolor=black,urlcolor=DarkBlue}
\renewcommand\d{\partial}
\newcommand\E{\mathbf{E}}
\newcommand\e{\mathbf{e}}
\newcommand\B{\mathbf{B}}
\renewcommand\b{\mathbf{b}}
\renewcommand\j{\mathbf{j}}

\newcommand\paral{\parallel}
\newcommand\ez{\mathbf{\hat z}}
\newcommand\<{\langle}
\renewcommand\>{\rangle}
\newcommand\sgn{\mathop{\mathrm{sgn}}}

\begin{document}
%\preprint{EFI-xx}
\title{Duality and universal transport in a mixed-dimension electrodynamics}

\author{Wei-Han Hsiao and Dam Thanh Son}
\affiliation{Kadanoff Center for Theoretical Physics, University of Chicago,
Chicago, Illinois 60637, USA}
\date{May 2017}

\begin{abstract}
We consider a theory of a two-component Dirac fermion localized on a
(2+1) dimensional brane coupled to a (3+1) dimensional bulk.  Using
the fermionic particle-vortex duality, we show that the theory has a
strong-weak duality that maps the coupling $e$ to $\tilde e=(8\pi)/e$.
We explore the theory at $e^2=8\pi$ where it is self-dual.  The
electrical conductivity of the theory is a constant independent of
frequency.  When the system is at finite density and magnetic field at
filling factor $\nu=\frac12$, the longitudinal and Hall conductivity
satisfies a semicircle law, and the ratio of the longitudinal and
Hall thermal electric coefficients is completely determined by the
Hall angle.  The thermal Hall conductivity is directly related to the
thermal electric coefficients.
\end{abstract}

\maketitle

\section{Introduction}

Recent developments have revealed powerful dualities between seemingly
different nonsupersymmetric quantum field theories in (2+1)
dimensions.  A special case, the bosonic particle-vortex duality, has
been known for decades~\cite{Peskin:1977kp,Dasgupta:1981zz}.  More
recently, a new duality between a free fermion theory and a gauge
theory called QED$_3$ was suggested in the context of the construction
of a particle-hole symmetric version of the composite fermion theory
of the half-filled Landau level~\cite{Son:2015xqa}.  Later, the
duality was shown to be related to the electromagnetic duality in
the bulk~\cite{Metlitski:2015eka,Wang:2015qmt}.  Most recently, the
duality was shown to be a particular case in a web of dualities that
follows from a relativistic version of flux
attachment~\cite{Karch:2016sxi,Seiberg:2016gmd}.  Many more examples
of dualities have been recently discovered, including those of
self-dual
theories~\cite{Xu:2015lxa,Cheng:2016pdn,Karch:2016aux,Metlitski:2016dht}.

In this paper we consider a simplest quantum theory living in mixed
(2+1) and (3+1) dimensions.  The theory involves a single
two-component Dirac fermion $\Psi$ living in a (2+1)D coupled to a massless
U(1) gauge field (photons) $A_\mu$ living in (3+1)D~\footnote{The Dirac fermion here should be understood as a domain wall fermion lying at the center of the entire bulk, rather than the boundary of half spacetime. The ``anomaly cancellation,'' in a sense, happens implicitly while integrating bulk fermion degrees of freedom. See Ref.~\onlinecite{Mulligan:2013he} for detailed discussion}
\begin{equation}\label{model}
  S = \int\!d^3x\, i\bar\Psi \gamma^\mu(\d_\mu - iA_\mu)\Psi
      - \frac1{4e^2}\int\!d^4x\, F_{\mu\nu}^2.
\end{equation}
This theory has been
previously considered in various
contexts~\cite{Kovner:1990zz,Dorey:1991kp,Marino:1992xi}.  Most
recently, it was considered in Ref.~\onlinecite{Son:2015xqa} as an example
of a relativistic theory exhibiting fractional quantum Hall effect.
The theory is similar to the low-energy effective theory describing
graphene---Dirac fermions in (2+1) dimensions interacting through a 3D
Coulomb potential---with two notable differences: there is only one
(instead of four) two-component Dirac fermion, and the photon
propagates with the same velocity as the fermion (instead of 300 times
faster as in graphene).

The bosonic version of the theory has been known for some
time~\cite{Fradkin:1996xb,Kuklov:2005,Geraedts:2012ut}.  In analogy
with the bosonic case, we find that the theory exhibits a strong-weak
duality, which combines the electromagnetic duality in the bulk and
the fermionic particle-vortex duality on the brane.  The duality maps
$e$ to $\tilde e=8\pi/e$, and the theory is self-dual at $e^2=8\pi$.
Provided that the theory is conformal at this coupling, we find
nontrivial consequences for the transport of the U(1) charge.  In
particular, we find that the electrical conductivity is equal to a
universal value:
\begin{equation}
  \sigma= \sigma_0 \equiv \frac1{4\pi} ~~ \left[ \sigma_0\equiv\frac{e^2}{2h} \right] .
\end{equation}
The expressions in the square brackets $[\cdots]$ on the right-hand
side in this and later equations correspond to the standard
normalization of current and gauge field, in which the electric charge
$e$ stays in the covariant derivative in Eq.~(\ref{model}).
Remarkably, the electrical conductivity is independent of the ratio
between the frequency and the temperature, $\omega/T$, and hence has
the same value in the ballistic ($\omega\gg T$) and hydrodynamic
($\omega\ll T$) regimes.  This behavior has previously been noted in
the strongly coupled large-$N$ theory living on a stack of
M2 branes~\cite{Herzog:2007ij}.

Moreover, we find that when one turns on a charge density $n$ and a
magnetic field $B$ satisfying the condition $n=B/(4\pi)$ (or filling
factor $\nu=\frac12$ in the quantum Hall terminology), there are nontrivial
relationships between electrical and thermal transport coefficients.
The longitudinal
and Hall conductivities satisfy a semicircle law:
\begin{equation}
  \sigma_{xx}^2 + \sigma_{xy}^2 = \sigma_0^2 .
\end{equation}
The ratio of the longitudinal and Hall thermoelectric coefficients is
directly related to the Hall angle
$\theta_H=\arctan(\sigma_{xy}/\sigma_{xx})$.  In addition, the thermal
Hall conductivity is related directly to the thermoelectric
coefficients.

The plan of this paper is as follows. We describe the model in
Sec.~\ref{sec:mixdimQED} and derive its self-duality in
Sec.~\ref{sec:derivation}.  In Sec.~\ref{sec:consequences} we extract
the consequences of the self-duality.  Section~\ref{sec:conclusion}
contains concluding remarks.

\section{Mixed-dimension QED}
\label{sec:mixdimQED}

We start to recall some feature of the
model~(\ref{model}) which we will call QED$_{4,3}$.
%This model can be
%defined to be gauge invariance and parity invariant at the same
%time~\cite{Mulligan:2013he}.

The coupling constant $e$ is dimensionless.  Physically, $e$
determines the force between two charges located infinitely far from
the brane, and hence it does not run.  The theory is scale invariant
at small $e$, but the situation at large $e$ is not clear.  The
large-$N$ version of~(\ref{model}) is conformal for all values of $e$,
including $e=\infty$ where the theory becomes $N$-flavor QED$_3$.  It
is expected that there exist a critical value $N_c$, below which
QED$_3$ undergoes spontaneous chiral symmetry breaking.  For $N<N_{\rm
  crit}$ then one expects QED$_{4,3}$ to be conformal only for
sufficiently small $e$, $e<e_{\rm crit}(N)$.  Analytic estimates for
$N_c$ based on a truncation of the Schwinger-Dyson equation are
typically of order 6--9~\cite{Gorbar:2001qt,Kotikov:2016yrn}, but a
recent numerical simulation~\cite{Karthik:2015sgq} suggests that scale
invariance persists at $N=2$, implying that $N_{\rm crit}<2$.  There
has not been any numerical study of $N=1$.  The ``strong version'' of
the conjectured duality between QED$_3$ and free Dirac
fermion~\cite{Metlitski:2015eka,Wang:2015qmt} would imply that
QED$_{4,3}$ is scale invariant even at $e^2=\infty$.

Applying the fermionic particle-vortex duality to the (2+1)D part
of~(\ref{model}), the model is mapped to
\begin{multline}\label{dual}
  S = \int\!d^3x\, \left[ i\bar\psi \gamma^\mu(\d_\mu - ia_\mu)\psi
   - \frac1{4\pi}\epsilon^{\mu\nu\lambda}A_\mu\d_\nu a_\lambda
  \right]\\
    - \frac1{4e^2}\int\!d^4x\, F_{\mu\nu}^2.
\end{multline}
where $\psi$ is the ``composite fermion,'' or the fermionic vortex.
Since $a_\mu$ is now a field propagating in (2+1)D, care is needed to
define the theory~(\ref{dual}) on a compact manifold---to avoid parity
anomaly one should either restrict the path integral over the $a$ field
configurations with even fluxes or introduce another gauge field
(for discussions of this point see
Refs.~\onlinecite{Karch:2016sxi,Seiberg:2016gmd}).  For the questions in which
we are interested in this paper, this subtlety will not play an important role.  At first sight~(\ref{dual}) and
(\ref{model}) appear to be very different theories; however we will
show that they are the same theory with different coupling constant.

%WHH edited

\section{Derivations of self-duality}
\label{sec:derivation}

\subsection {A simple derivation}

The most straightforward way to see the
self-duality is to rewrite both theories in the form of theories with
nonlocal current-current interactions.  Integrating over $A_\mu$ in
Eq.~(\ref{model}) we obtain a nonlocal action in $2+1$ dimensions,
\begin{equation}\label{nonlocal}
  S= \!\int\!d^3x\, i\bar\Psi\gamma^\mu\d_\mu\Psi -i\frac{e^2}{2}\!\int\! d^3x\, d^3x'\, j_{\Psi\mu}\frac{1}{\sqrt{\d^2}}j_{\Psi}^{\mu}.
\end{equation}
where $\frac{1}{\sqrt{\partial^2}}$ is the (3+1)D Feynman propagator subject to the constraint $z=z'=0$.
On the other hand, integrating over $A_\mu$ in the dual
theory~(\ref{dual}) leads to
\begin{equation}\label{nonlocal-dual1}
S= \!\int\! d^3x\, i\bar{\psi}\gamma^\mu(\d_{\mu}-ia_{\mu})\psi-\frac{ie^2}{(8\pi)^2}\!\int\!d^3x\, d^3x'\, f_{\mu\nu}\frac{1}{\sqrt{\d^2}}f^{\mu\nu},
\end{equation}
where $f_{\mu\nu}=\d_{\mu}a_{\nu}-\d_{\nu}a_{\mu}$, and now integrating over $a_\mu$ we get
\begin{equation}\label{nonlocal-dual2}
 S = \!\int\!d^3x\, i\bar\psi\gamma^\mu\d_\mu\psi -i\frac{(8\pi)^2}{2e^2}\!\int\! d^3x\, d^3x'\, j_{\psi\mu}\frac{1}{\sqrt{\d^2}}j_{\psi}^{\mu},
\end{equation}
which has the same form as Eq.~(\ref{nonlocal}), with the replacement
$e\to 8\pi/e$.

\subsection{Alternative derivation through bulk electromagnetic duality}

There is another derivation of the self-duality which reveals the
connection to electromagnetic duality in the bulk.  We first note that
across the brane, the density and current on the brane determines the
jump of the perpendicular (to the brane) component of the electric
field and the parallel components of the magnetic field,
\begin{equation}\label{jump}
  \Delta E_z = e^2\rho, \qquad
  \Delta \B_\paral = e^2\,\j \times \ez.
\end{equation}
% WHH edited.  
In contrast, $B_z$ and the
$\E_\paral$ are continuous across the brane.

Without losing generality, we can impose an orbifold condition 
\begin{subequations}\label{A-orb}
\begin{align}
  A_\mu(z) &= A_\mu(-z), \qquad \alpha=t,x,y\\
  A_z(z) &= -A_z(-z).
\end{align}
\end{subequations}
One can see that by decomposing the fields into symmetric and
antisymmetric (under $z\to-z$) parts, $A_\mu=A_\mu^s+A_\mu^a$, the
action then decomposes into
\begin{equation}
  S[\psi, A_\mu^s, A_z^a] + S[A_\mu^a, A_z^s].
\end{equation}
The fields $A_\mu^a$ and $A_z^s$ do not couple to the brane degrees
of freedom and can be integrated away.  With the orbifold
condition~(\ref{A-orb}), Eqs.~(\ref{jump}) completely determine the
boundary values of the perpendicular component of the electric field
and the parallel components of the magnetic field,
\begin{subequations}\label{orb_bc}
\begin{align}
  E_z(z=\pm\epsilon) &= \pm \frac12 e^2\rho ,\\
  \B_\paral(z=\pm\epsilon) &= \pm \frac12 e^2 \j \times \ez.
\end{align}
\end{subequations}
In contrast $\E_\paral$ and $B_z$ are continuous at $z=0$.

We now analyze the composite fermion theory~(\ref{dual}).  First let
us write down the field equations.  $A_\mu$ satisfy the Maxwell
equation in the bulk and the boundary conditions~(\ref{orb_bc}), where
the charge density and current are
\begin{subequations}\label{source_cf}
\begin{align}
  \rho &= - \frac1{4\pi} b , \\
  \j &= -\frac1{4\pi} \e \times \ez \label{je}.
\end{align}
\end{subequations}
Varying $S$ with respect to $a_\mu$ we also find, at $z=0$,
\begin{subequations}\label{jCF}
\begin{align}
  \rho^{\rm CF} &= \frac1{4\pi} B_z , \\
  \j^{\rm CF} &= \frac1{4\pi} \E_\paral\times\ez.
\end{align}
\end{subequations}

Instead of dealing with $A_\mu$, we now perform an operation
electromagnetic duality in the bulk.  We introduce a dual
electromagnetic field $\tilde\E$ and $\tilde\B$ related to $\E$ and
$\B$ by
\begin{subequations}\label{EMduality}
\begin{align}
  \E &=  -\sgn(z)\tilde\B , \\
  \B &= \sgn(z)\tilde \E .
\end{align}
\end{subequations}
Note that the transformation has a discontinuity at $z=0$.  In the
bulk $\tilde\E$ and $\tilde\B$ satisfy the same free Maxwell equations
as $\E$ and $\B$.  Note that the orbifold conditions~(\ref{A-orb}) are
preserved by the duality transformation~(\ref{EMduality}).

With Eq.~(\ref{source_cf}), Eqs.~(\ref{orb_bc}) become, after the EM
duality,
\begin{subequations}
\begin{align}
  \tilde B_z &= - \frac{e^2}2 \rho = \frac{e^2}{8\pi}b , \\
  \tilde \E_\paral &= \frac{e^2}2 \j\times\ez = \frac{e^2}{8\pi} \e.
\end{align}
\end{subequations}
That means we can now extend the gauge field $a_\mu$ to the whole
(3+1)D space, taking for the value of the field in the bulk
$a_\mu=\frac{8\pi}{e^2}\tilde A_\mu$.  The bulk Lagrangian for
$a_\mu$ is
\begin{equation}
  S_{\rm bulk}[a] = -\frac1{4e^2} \left(\frac{e^2}{8\pi}\right)^2
    f_{\mu\nu}^2 = - \frac1{4\tilde e^2} f_{\mu\nu}^2 ,
\end{equation}
where $\tilde e = {8\pi}/e$.
The electromagnetic duality operation~(\ref{EMduality}) introduces
jumps in components of $f_{\mu\nu}$,
\begin{subequations}
\begin{align}
  e_z (z=\pm\epsilon) &= \frac{8\pi}{e^2} \tilde E_z(z=\pm\epsilon) =
  \pm \frac{8\pi}{e^2} B_z ,\\
  \b_\paral(z=\pm\epsilon) &=\frac{8\pi}{e^2}\tilde \B_\paral(z=\pm\epsilon)
  = \mp \frac{8\pi}{e^2}\E_\paral .
\end{align}
\end{subequations}
By using Eq.~(\ref{jCF}), these equations can be written as
\begin{subequations}
\begin{align}
  e_z(z=\pm\epsilon) &= \pm 4\pi \frac{8\pi}{e^2}\rho_{\rm CF}
  = \pm \frac12 \tilde e^2 \rho_{\rm CF} , \\
  \b_\paral(z=\pm\epsilon) &= \pm 4\pi\frac{8\pi}{e^2} \j_{\rm CF}\times\ez
  = \pm \frac12 \tilde e^2 \j_{\rm CF} \times \ez ,
\end{align}
\end{subequations}
which have exactly the same form as Eq.~(\ref{orb_bc}).

Thus the action for the composite fermion can be written as
\begin{equation}
  S = \int\!d^3x\, i\bar\psi \gamma^\mu(\d_\mu - ia_\mu)\psi
      - \frac1{4\tilde e^2}\int\!d^4x\, f_{\mu\nu}^2 .
\end{equation}

When $e^2=8\pi$ this action coincides with the action for the original
electron.  This is the self-dual point.

\subsection{Comparison with the model in half-space}

In the literature, one frequently considers a model where the gauge
field propagates in one half space and the fermion is localized on
the boundary of the half space.  In this case (see, e.g.,
Ref.~\onlinecite{Seiberg:2016gmd}) we need a bulk $\theta$ term with
$\theta=\pi$ to properly define the partition function.  The duality
transformation considered above becomes a combination of $\mathsf S$
and $\mathsf T$ transformations.  To see that, let us define the
complex coupling constant
\begin{equation}
  \tau = \frac{\theta}{2\pi}+i \frac{2\pi}{e^2}
\end{equation}
and recall that the operations $\mathsf S$ and $\mathsf T$ act on the
constant as
\begin{align}
  \mathsf S: & ~ \tau\to -1/\tau\\
  \mathsf T: & ~ \tau\to \tau+1.
\end{align}
The composite operator $\mathsf S\mathsf T^{-2}\mathsf S\mathsf
T^{-1}$ maps $\tau$ onto
\begin{align}
\tau \to \tau' = \frac{\tau-1}{2\tau-1}.
\end{align}
In particular, starting from $\theta = \pi$, one also ends up with
$\theta'=\pi$:
\begin{align}
\tau = \frac{1}{2}+i \frac{2\pi}{e^2}\to \tau' = \frac{1}{2}+i \frac{e^2}{4\cdot 2\pi}.
\end{align}
The self-dual point is at $e^2 = 4\pi$.  This is twice smaller than
the value $e^2=8\pi$ we found for the model living in the whole space.
The factor of 2 difference accounts for the fact that in our current
model the electric field lines are restricted to one half space;
hence the strength of the Coulomb interaction is twice larger than in
the model living in the whole space with the same value of $e^2$.

\section{Consequences of self-duality}
\label{sec:consequences}

\subsection{Electrical conductivity at zero chemical potential and zero magnetic field}

Now let us explore the consequences of duality for the conductivity.
Let us introduce the conductivity tensor, which is denoted as
$\sigma_{ij}$ on the electron side and $\tilde\sigma_{ij}$ on the
composite fermion side.

On the electron side Ohm's law reads
\begin{equation}
  j^i = \sigma^{ij} E_j ,
\end{equation}
and on the composite fermion side
\begin{equation}
  j_{\rm CF}^i = \tilde \sigma^{ij} e_j.
\end{equation}
Using the duality dictionary,
\begin{align}
  j^i &= -\frac1{4\pi}\epsilon^{ij} e_j , \\
  j^i_{\rm CF} & = \frac1{4\pi}\epsilon^{ij} E_j ,
\end{align}
one can easily find
\begin{equation}\label{dualRelation}
  \sigma  = - \frac1{(4\pi)^2} \epsilon\, \tilde\sigma^{-1}\epsilon,\quad
  \epsilon = \begin{pmatrix} 0 & 1 \\ -1 & 0 \end{pmatrix}.
\end{equation}

Let us assume the electron theory to be at zero chemical
potential and zero magnetic field, but finite temperature.  The
conductivity tensor is diagonal: $\sigma_{ij}=\sigma\delta_{ij}$.
Equation~(\ref{dualRelation}) now implies
\begin{comment}
\begin{equation}
  \j_\paral = \sigma \E_\paral, \qquad \sigma=\sigma(e).
\end{equation}
On the other hand, on the CF side, Ohm's law reads
\begin{equation}\label{OhmCF}
  \j^{\rm CF}_\paral = \tilde \sigma \e_\paral, \qquad
  \tilde \sigma = \sigma(\tilde e).
\end{equation}
From the CF effective action we have Eq.~(\ref{je}) and 
\begin{align}
  \j^{\rm CF}_\paral = \frac1{4\pi} \E_\paral\times \ez,
\end{align}
which allow us to rewrite Eq.~(\ref{OhmCF}) as
\begin{equation}
  \j_\paral = \frac1{(4\pi)^2\tilde\sigma}\E_\paral.
\end{equation}
This means
\end{comment}
\begin{equation}\label{sstild}
  \sigma(e)\tilde{\sigma}(\tilde e) = \frac1{(4\pi)^2}.
\end{equation}
In particular, at the self-dual point $e^2=\tilde e=8\pi$,
\begin{equation}\label{sigma-selfdual}
  \sigma = \sigma_0.
\end{equation}

At zero temperature, the conductivity
is just the coefficient appearing in the current-current correlation
function:
\begin{equation}
  \< J_\alpha(p) J_\beta(q)\> = \frac{\sigma}{\sqrt{q^2}} (q^2 g_{\mu\nu}-q_\mu q_\nu)
  (2\pi)^3\delta^{(3)}(p+q).
\end{equation}
For a free Dirac fermion, $e^2=0$, $\sigma=1/16$~\cite{LFSG}.  Our
result thus indicates that the conductivity at $e^2=8\pi$ is by a
factor of $4/\pi\approx1.273$ times larger than at $e^2=0$.  This can
be compared with the weak-coupling result derived in
Ref.~\onlinecite{Teber:2016unz},
\begin{equation}
  \sigma = \frac1{16}\left( 1+ C
     \frac{e^2}{4\pi} +O(e^4) \right), \qquad  C = \frac{92-9\pi^2}{18\pi}\,.
\end{equation}
If one naively substitutes $e^2=8\pi$, one finds that the one-loop
correction enhances the conductivity by a factor of $1.112$.
Obviously at such large coupling higher-loop effects cannot be
neglected.

From Eq.~(\ref{sstild}) we also find the conductivity at infinite
coupling,
\begin{equation}
  \sigma(e^2=\infty) = \frac1{\pi^2} ~~~ \left[ \frac2\pi \frac{e^2}h \right].
\end{equation}

Moreover, our result~(\ref{sigma-selfdual}) is also applicable at
finite temperature, where it implies that the conductivity
is independent of frequency.  In particular the conductivity has the
same value in the hydrodynamic regime $\omega\ll T$ and in the
ballistic regime $\omega\gg T$.  A frequency-independent
finite-temperature conductivity has been found in
Ref.~\onlinecite{Herzog:2007ij} for the theory living on a stack of $N$
M2 branes in M theory in the limit of large $N$, which has been traced
back to the electromagnetic duality in the holographic description.

The result continues to be true in the presence of duality symmetric
disorder.  Such disorder can be introduced, e.g., as a randomly
fluctuating mass term of the fermion, or by placing random electric
charges and magnetic monopoles in the bulk near the brane so that the
statistical properties of this random ensemble of electric and
magnetic charges is invariant under electromagnetic duality.

One can also consider transport at nonzero wave vectors, where it is
characterized by the longitudinal and transverse conductivities.
Again from Eq.~(\ref{dualRelation}) it follows that at $e^2=8\pi$,
\begin{equation}
  \sigma_\perp(\omega,q)\sigma_\paral(\omega,q) = \sigma_0^2\,.
\end{equation}
This exact relationship has been found previously in the context of
holography~\cite{Herzog:2007ij} and the bosonic self-dual
theory~\cite{Geraedts:2012ut}. When $q=0$ the longitudinal and
transverse conductivities are equal and one recovers
Eq.~(\ref{sigma-selfdual}).

\subsection{Electric and thermal transport at filling factor $\nu=\frac12$}

\subsubsection{Electrical conductivities}

We now consider our system in a finite magnetic field and
finite density, so that the filling factor is $1/2$.  At zero
temperature and at weak coupling, the system forms an integer quantum Hall
state with the zero-energy Landau level completely filled.
Now the conductivity tensor has nonzero off-diagonal elements (the Hall
conductivity).  For simplicity we only consider transport at zero
wave number when
\begin{equation}
  \sigma_{ij} = \begin{pmatrix} \sigma_{xx} & \sigma_{xy} \\
  - \sigma_{xy} & \sigma_{xx} \end{pmatrix}.
\end{equation}
From the duality mapping between the density and the magnetic fields,
it follows that at the self-dual point the dual theory is exactly the
original theory, but with the filling factor $-\frac12$.  That means
$\tilde\sigma_{xx}=\sigma_{xx}$, $\tilde\sigma_{xy}=-\sigma_{xy}$, or
$\tilde\sigma=\sigma^T$, and Eq.~(\ref{dualRelation}) implies
\begin{equation}\label{semicircle}
  \sigma_{xx}^2 + \sigma_{xy}^2 = \sigma_0^2.
  %\frac1{(4\pi)^2}.
\end{equation}
At zero temperature and in the absence of disorder, we have an integer
quantum Hall state with $\sigma_{xx}=0$ and $\sigma_{xy}=\frac1{4\pi}$
and Eq.~(\ref{semicircle}) is trivially satisfied.  Turning on the
temperature, a nonzero $\sigma_{xx}$ is induced by scatterings of the
charge fermions on photons in the bulk.  In the limit when the
temperature is very large (compared to the scale set by the magnetic
field and the density), $\sigma_{xx}=\frac1{4\pi}$ as we have derived
at zero field and zero chemical potential.  As the temperature changes
the conductivities vary between these two extreme, following a
quarter circle in the ($\sigma_{xx}$, $\sigma_{xy}$) plane.  This
behavior is reminiscent of the semicircle law in quantum Hall
transitions~\cite{Dykhne:1993yn,Ruzin:1995,Burgess:2000kj}.

\subsubsection{Thermoelectric coefficients}

We now apply the duality
technique to the thermoelectric transport.  Introducing the
thermoelectric coefficients $\alpha_{xx}$ and $\alpha_{xy}$,
\begin{align}
  j^i = \sigma^{ij} E_j + \alpha^{ij} \d_j T,\\
  j_{\rm CF}^i = \tilde \sigma^{ij} e_j + \tilde\alpha^{ij} \d_j T.
\end{align}
By using again the duality mapping, we find
\begin{equation}\label{alpha-dual}
  \alpha = \frac1{4\pi} \epsilon\tilde\sigma^{-1} \tilde\alpha.
\end{equation}
Again, at the self-dual point $\tilde\alpha=\alpha^T$, and
Eq.~(\ref{alpha-dual}) determines the ratio $\alpha_{xy}/\alpha_{xx}$
in terms of $\sigma_{xy}/\sigma_{xx}$.  If we introduce the Hall angle
$\theta_H=\arctan(\sigma_{xy}/\sigma_{xx})$,
%\begin{equation}
%  \sigma_{xy} = \frac1{4\pi}\cos\theta,\qquad
  %  \sigma_{xx} = \frac1{4\pi}\sin\theta,
%  \sigma_{xy} = \frac1{4\pi}\sin\theta_H,\qquad
%  \sigma_{xx} = \frac1{4\pi}\cos\theta_H, 
%\end{equation}
then
\begin{equation}
%  \frac{\alpha_{xx}}{\alpha_{xy}} = \tan\frac\theta2
%  = \left( 1+\frac{\sigma_{xy}^2}{\sigma_{xx}^2}\right)^{1/2}
%    - \frac{\sigma_{xy}}{\sigma_{xx}}.
  \frac{\alpha_{xy}}{\alpha_{xx}} = \tan\left(\frac\pi4+\frac{\theta_H}2\right).
%  = \left( 1+\frac{\sigma_{xy}^2}{\sigma_{xx}^2}\right)^{1/2}
%    - \frac{\sigma_{xy}}{\sigma_{xx}}.
\end{equation}
This can be written in terms of components of the Seebeck tensor,
defined through $E_i=S_{ij}\d_jT$ when there is no current, $j_i=0$,
\begin{equation}
  \frac{S_{xy}}{S_{xx}} = \tan\left(\frac\pi4-\frac{\theta_H}2\right).
\end{equation}
Note that $S_{xx}$ is the usual Seebeck coefficient and $S_{xy}/B$ is
the Nernst coefficient.

Again one can discuss two limits.  In the low-temperature quantum Hall
regime, $\theta\to \pi/2$, the Hall thermoelectric coefficient
$\alpha_{xy}$ dominates over the longitudinal coefficient
$\alpha_{xx}$.  In the very high temperature regime, $T\gg n^{1/2}$,
our result indicates that the ratio $\alpha_{xx}/\alpha_{yy}$ tends to
1, in contrast to the electric conductivities where $\sigma_{xx}$
dominates over $\sigma_{xy}$.  This is not completely surprising: the
thermoelectric coefficients break charge conjugation and hence are zero
when $n=B=0$, and one can show that $\alpha_{xx}$ is proportional to
$n$ and $\alpha_{xy}$ to $B$ when $n$ and $B$ are small.  At filling
factor $\nu=1/2$, $n$ and $B$ are proportional to each other; hence
$\alpha_{xx}$ and $\alpha_{xy}$ are of the same order of magnitude.

\subsubsection{Thermal Hall coefficient}

Finally, we consider the transport of heat at filling factor $\nu=\frac12$.
The heat current is
\begin{equation}
q_i = - T\alpha_{ij}E_j - \bar\kappa_{ij}\partial_jT,
\end{equation}
where $\bar\kappa_{ij}$ is the thermal conductivity tensor in the absence
of electric field, which is related to the thermal conductivity tensor in the absence
of electric current $\kappa_{ij}$ by $\kappa=\bar\kappa-T\alpha\sigma^{-1}\alpha$.  As the heat current is invariant
under electromagnetic duality, it has to be given by the same
expression in the dual description,
\begin{equation}
q_i = - T\tilde \alpha_{ij} e_j - \tilde{\bar\kappa}_{ij}\partial_jT .
\end{equation}
Following the duality maps we obtain a connection relating the thermal
conductivity tensors on the two sides the duality,
\begin{align}\label{kktilde}
  \bar\kappa = \tilde{\bar\kappa}- T\tilde{\alpha}\tilde{\sigma}^{-1}\tilde{\alpha} = \tilde\kappa.
\end{align}
Thus, the thermal conductivity at zero field on one side of the duality is equal to the
thermal conductivity at zero current on the other side.
At the self-dual point, $\tilde\alpha=\alpha^T$ and
$\tilde{\bar\kappa}=\bar\kappa^T$.  Equation~(\ref{kktilde}) then establishes a
direct relationship between the thermal Hall conductivity and the
thermoelectric coefficients:
\begin{equation}
  \bar\kappa_{xy} = -\kappa_{xy} = \frac T2 \frac{\alpha_{xx}^2+\alpha_{xy}^2}
        {\sigma_0}
  =\frac T2 (S_{xx}^2+S_{xy}^2) \sigma_0 \,.
%   ~~~ \left[ \frac {hk_BT}{e^2} (\alpha_{xx}^2+\alpha_{xy}^2) \right]
\end{equation}
We also find that $\bar\kappa_{xx}=\kappa_{xx}$, but otherwise there is 
no constraint on this coefficient.  Not surprisingly, the values
of the kinetic coefficients in the $N$ M2-brane
theory~\cite{Hartnoll:2007ih} also respects an analogous constraint
(in the dc regime).  We note that relationships similar to
Eqs.~(\ref{alpha-dual}) and (\ref{kktilde}) have been found in
Ref.~\onlinecite{Donos:2017mhp} in the context of a holographic model with
bulk electromagnetic duality.

\section{Conclusion}
\label{sec:conclusion}

We have shown that the simple model of a (2+1)D fermion coupled to a
three-dimensional U(1) gauge field, QED$_{4,3}$, exhibits weak-strong
duality, and is self-dual at a particular value of the coupling
constant.  From the self-duality we derive the value of the
conductivity at zero density and magnetic field and show that it is
independent of frequency.  At finite magnetic field and filling factor
$\nu=\frac12$ we were able to derive a semicircle law satisfied by the
longitudinal and Hall conductivities, relate the ratio of the diagonal
and Hall thermoelectric coefficients with the Hall angle, and derive a
relationship between the thermal Hall conductivity and the
thermoelectric coefficients.

We note here that all results obtained above can be transferred to the
bosonic model, at zero density and magnetic field and at filling
factor $\nu=1$.  This may be interesting since the $\nu=1$ bosonic
quantum Hall state can be a Fermi liquid.  Most formulas derived in
the paper remain valid if one replaces the value of $\sigma_0$ by
$\sigma_0=1/(2\pi)$.
%Besides, since they originate from the S-duality
%in the 3+1 D spacetime,

It remains to be seen whether self-dual systems are accessible
experimentally~\cite{Breznay:2016}.  If they are, the predictions made
in this paper may serve as tests of self-duality in such systems.

\emph{Note added.}---%
%During the completion of this manuscript
Recently, we
became aware of a work by D.~F.~Mross, J.~Alicea, and O.~I.~Motrunich
\cite{Mross:2017gny} which partially overlaps with our paper. We thank them for
sharing their unpublished work with us.

\acknowledgments

The authors thank Clay C\'ordova, Olexei Motrunich, Cenke Xu, Luis Melgar, Subir Sachdev, and Nathan Seiberg for
discussions.  This work is supported, in part, by U.S.\ DOE Grant
No.\ DE-FG02-13ER41958
%the ARO MURI grant No.\ 63834-PH-MUR,
and a Simons Investigator Grant from the Simons Foundation.  Additional
support was provided by the Chicago MRSEC, which is funded by NSF
through Grant No.\ DMR-1420709.

\bibliography{selfdual-final}

\end{document}